\documentclass[]{article} 
\usepackage{emulateapj}
\usepackage{graphicx}
\usepackage{multirow}

\advance \voffset by -0.5cm\relax

\usepackage{color}

\def\msun{{\rm ~M}_{\odot}}
\def\rsun{{\rm ~R}_{\odot}}

\def\mpy{{\rm ~M}_{\odot} {\rm ~yr}^{-1}}

\begin{document}

\title{Cyg X-3: a Galactic double black hole or black hole-neutron star progenitor}

 \author{Krzysztof Belczynski\altaffilmark{1,2},
         Tomasz Bulik\altaffilmark{1}, 
         Ilya Mandel\altaffilmark{3}, 
         B.S. Sathyaprakash\altaffilmark{4},
         Andrzej Zdziarski\altaffilmark{5}, 
         Joanna Mikolajewska\altaffilmark{5}} 

 \affil{
     $^{1}$ Astronomical Observatory, University of Warsaw, Al.
            Ujazdowskie 4, 00-478 Warsaw, Poland\\
     $^{2}$ Center for Gravitational Wave Astronomy, University of Texas at
            Brownsville, Brownsville, TX 78520, USA\\
     $^{3}$ School of Physics and Astronomy, University of Birmingham,
            Edgbaston, B15 2TT, UK\\
     $^{4}$ School of Physics and Astronomy, Cardiff University, 5, The
            Parade, Cardiff, UK, CF24 3YB\\
     $^{5}$ Nicolaus Copernicus Astronomical Center, PAN, ul. Bartycka 18,
            00-716 Warsaw, Poland\\ 
 }
 
\begin{abstract}
There are no known double black hole (BH-BH) or black hole-neutron star (BH-NS)
systems. We argue that Cyg X-3 is a very likely BH-BH or BH-NS progenitor. This 
Galactic X-ray binary consists of a compact object, wind-fed by a Wolf-Rayet 
(WR) type companion. Based on a comprehensive analysis of observational data, it was 
recently argued that Cyg X-3 harbors a $2$--$4.5 \msun$ black hole (BH) and a 
$7.5$--$14.2 \msun$ WR companion. We find that the fate of such a binary leads 
to the prompt ($\lesssim 1$ Myr) formation of a close BH-BH system for the high 
end of the allowed WR mass ($M_{\rm WR} \gtrsim 13 \msun$). For the low- to 
mid-mass range of the WR star ($M_{\rm WR} \sim 7$--$10 \msun$) Cyg X-3 is most likely 
(probability $70\%$) disrupted when WR ends up as a supernova. However, with 
smaller probability, it may form a wide ($15\%$) or a close ($15\%$) BH-NS 
system. The advanced LIGO/VIRGO detection rate for mergers of BH-BH systems from 
the Cyg X-3 formation channel is $\sim 10$ yr$^{-1}$, while it drops down to 
$\sim 0.1$ yr$^{-1}$ for BH-NS systems. If Cyg X-3 {\em in fact} hosts a low 
mass black hole and massive WR star, it lends additional support for the 
existence of BH-BH/BH-NS systems. 
\end{abstract}


\section{Introduction}

Ever since the discovery of the Hulse-Taylor pulsar 1913+16 (Weisberg \&  Taylor 2005), 
a system consisting of a pair of neutron stars, there has been a lot of effort to 
detect other double compact objects. So far only double neutron star (NS-NS) systems
were found (e.g., Lorimer 2008). In particular, no BH-NS nor BH-BH systems are known 
at the moment. 

Merging compact binaries have provided the strongest motivation yet to build wide-band
interferometric gravitational wave detectors.  Second generation detectors, the Laser 
Interferometer Gravitational-wave Observatory (LIGO) in the US (Harry, 2010), 
Virgo in Italy (Acernese et al., 2006) and KAGRA (KAGRA 2012) in Japan, have the potential to 
observe the late inspiral and merger phases of compact binaries with their total mass 
in the range $\sim [1,\,10^3]\,M_\odot.$ The range of masses they could observe is 
determined by the frequency window of $\sim[10,10^4]\,\rm Hz$ in which they operate. 
Radio binary pulsars lend strong motivation for systems at the lower end of the above 
mass range. When LIGO, Virgo and KAGRA begin their observations later in this decade, 
they could well provide the observational evidence for BH-NS/BH-BH systems or set 
very strong constraints on the statistics of their populations. However, we now have 
a plausible progenitor of a BH-NS/BH-BH system in our own galaxy.

High mass X-ray binaries (HMXBs) provide a unique opportunity to study various
astrophysical phenomena. At {\em Warsaw Observatory} we have undertaken a
program to study HMXBs in the context of the formation of double compact objects 
like BH-BH or BH-NS systems. 
We have already provided studies of two extragalactic HMXBs that host the
most massive known BHs of stellar origin: IC10 X-1 and NGC 300 X-1 (Bulik,
Belczynski \& Prestwich 2011), an analysis of Cyg X-1 binary that harbors the most 
massive stellar BH in our Galaxy (Belczynski, Bulik \& Bailyn 2011) and 
analyses of several other binaries with well established parameters: GX 301-2, 
Vela X-1, XTEJ1855-026, 4U1907+09, Cir X-1, LMC X-1, LMC X-3, M33 X-1 
(Belczynski, Bulik \& Fryer 2012b). 

In this study, we follow the recent estimate of system parameters for another 
Galactic HMXB: Cyg X-3. Zdziarski, Mikolajewska \& Belczynski (2012) have 
estimated that Cyg X-3 consists of a low mass BH, $2$--$4.5\msun$, and
a massive WR star, $7.5$--$14.2\msun$. We use this estimate to
calculate the future evolution of Cyg X-3 to check whether this binary may
provide any observational constraints on as-yet undetected BH-BH and BH-NS 
systems. The past evolution of Cyg X-3 was studied in detail by Lommen et
al. (2005).

\section{Estimates}

\subsection{The future evolution of Cyg X-3}

In this section we describe in detail Cyg X-3 evolutionary scenarios leading 
to the formation of a BH-BH or BH-NS system. For demonstration we choose
extremes of the allowed mass range of BH and WR components: 
$M_{\rm BH1}=2.0$ with $M_{\rm WR}=7.5 \msun$ and $M_{\rm BH1}=4.5$ with 
$M_{\rm WR}=14.2\msun$.
Additionally, we evolve the most likely configuration $M_{\rm BH1}=2.4$ with 
$M_{\rm WR}=10.3 \msun$ as adopted from Zdziarski et al. (2012).
First, we employ our standard model for stellar and binary evolution (see below)
and in the next section we will test how various evolutionary uncertainties
change our results. 

We evolve three $M_{\rm zams}=27, 36, 50\msun$ single stars (Hurley, Pols \& Tout 
2000) at solar metallicity $Z=0.02$ with our calibrated wind mass loss rates 
(Belczynski et al. 2010a). At $t=6.5, 5.2, 4.4$ Myr these stars leave the main sequence 
(MS) and become Hertzsprung gap objects with helium core of $M_{\rm WR,i}=7.5, 10.3, 
14.2 \msun$, respectively. 
We expose this core right after MS and make them naked helium stars: massive WR
objects. We pair these WR stars with $M_{\rm BH1} = 2.0, 2.4, 4.5 \msun$ BHs
and create three binaries all with orbital period of $P_{\rm orb}=4.8$hr as
observed for Cyg X-3. 
Such an approach allows a WR star to shed a maximum amount of mass, reducing
its chance to become NS or BH, and thus makes our estimates conservative. We 
should note that this simplified picture in which the WR star is evolved 
independently of its companion ignores the effect of past interactions with the 
BH companion on the future evolution of the WR star, which could change the 
rate of future mass loss. However, the mass loss rates that we apply to WR
companion are derived based on a large population of WR stars, including those in binaries
(see below). 

The binary separations at the currently observed orbital period of the three
synthetic binaries are $a=3.0, 3.4, 3.8\rsun$ and the Roche lobe radii of WR 
component are $R_{\rm lobe}= 1.5, 1.7, 1.8\rsun$. The corresponding radii of 
WR stars are $R_{\rm WR}=0.8, 1.0, 1.2\rsun$. Massive helium stars do not 
expand at any significant level during evolution. Therefore, no Roche lobe 
mass transfer episode is expected in the future evolution of these systems. 

The lifetimes of these WR stars are $t_{\rm life}=0.95, 0.77, 0.64$ Myr and 
at the end of their evolution the masses drop to $M_{\rm WR,f}=5.5, 6.8, 8.2 \msun$. 
We have applied wind mass loss rates adopted from Hamann \& Koesterke (1998) 
that take into account effects of clumping 
\begin{equation}
(dM/dt) = 10^{-13} \left({L \over L_\odot}\right)^{1.5} \mpy.  
\label{theory}
\end{equation}
This prescription is based on detailed stellar evolutionary calculations 
combined with comprehensive WR wind models and calibrated on observations 
of three WR stars. We obtain WR star luminosity ($L$) from evolutionary
models of Hurley et al. (2000).

At the time of explosion, orbits have expanded to $a=3.9, 4.7, 5.6\rsun$ 
due to the WR wind mass loss and WR stars have formed massive CO cores 
($M_{\rm CO}=4.0, 5.0, 6.2\msun$).
The WR stars undergo a core collapse. We use Fryer et al. (2012) rapid 
explosion model that reproduces the observed mass gap between neutron stars 
and black holes (Belczynski et al. 2012a). 
For initial $M_{\rm WR,i}=7.5, 10.3 \msun$ WR stars the core collapse is followed by 
Type Ib supernova and neutron stars form with mass $M_{\rm NS}=1.5, 1.7\msun$, 
respectively. The mass loss and natal kicks associated with supernova are
very likely to disrupt these two synthetic binaries ($f_{\rm disruption}=0.69$). 
However, there is also a significant chance to form a close BH-NS system. We 
estimate the probability of the BH-NS system formation with gravitational merger 
time $T_{\rm merger}$ shorter than $10$ Gyr at the level $f_{\rm close}=0.14, 0.12$ 
for the low and intermediate mass realization of Cyg X-3. We have applied a 
distribution of natal kicks as derived from the population of Galactic pulsars 
(Hobbs et al. 2005). The kicks have a random direction and their magnitude is 
taken from a single Maxwellian with $\sigma=265$ km s$^{-1}$. Additionally, 
we lower the magnitude of natal kicks due to the amount of fall back expected 
in each supernova explosion $V_{\rm kick}=(1-f_{\rm fb}) V_{\rm kick}$ (Fryer et al. 
2012).  The fall-back amount, the fraction of matter that was initially ejected in 
a supernova explosion but that is later accreted back onto the compact object, was 
estimated to be $f_{\rm fb}=0.14, 0.16$ in case of models with initial 
$M_{\rm WR,i}=7.5, 10.3 \msun$, respectively. 

For the high component-mass realization ($M_{\rm WR,i}=14.2 \msun$), within our 
rapid supernova explosion model the entire WR star falls into a BH (i.e., 
$f_{\rm fb}=1.0$). We assume the loss of $10\%$ of the 
gravitational mass during the collapse through neutrino emission, which leads to a slight
orbital expansion and a small induced eccentricity ($a=6.0\rsun$, $e=0.07$). In 
the end the second BH forms with mass $M_{\rm BH,2}=7.4\msun$. No natal kick is
imparted on a BH within this model.
Therefore, a close BH-BH system forms with a chirp mass of $M_{\rm c,dco} \equiv 
(M_1 M_2)^{3/5} (M_1 + M_2)^{-1/5} =5.0\msun$ and a merger time of 
$T_{\rm merger}=0.5$ Gyr. 
In this particular case with no natal kick there is no other possibility 
($f_{\rm close}=1.0$), within the framework of our model, then to form a close 
($T_{\rm merger}<10$ Gyr) BH-BH system.

If we conservatively assume that at any given time 
there is only $1$ such system as Cyg X-3 it means that Galactic birth rate is 
at the level ${\cal R}_{\rm birth}=1/t_{\rm WR}$. As listed above, lifetimes of 
the considered massive WR stars are very short $t_{\rm WR} \lesssim 1$ Myr. 
Since the merger times of systems that we include in our analysis are relatively
short ($T_{\rm merger}<10$ Gyr) and the star formation was approximately constant 
in the Galactic disk over a long period of time ($\sim 10$ Gyr) the Galactic 
merger rate of BH-NS and BH-BH systems may be estimated from 
\begin{equation} \label{eq:rate}
{\cal R}_{\rm MW} = f_{\rm close} {\cal R}_{\rm birth} \sim f_{\rm close} \ {1 \over
t_{\rm WR}};
\end{equation}
we discuss statistical uncertainties in the estimate of the birth rates in section 2.4.
For our three Cyg X-3 realizations we obtain ${\cal R}_{\rm MW}=0.19, 0.20$ 
Myr$^{-1}$ in the cases of BH-NS formation and ${\cal R}_{\rm MW}=1.56$ 
Myr$^{-1}$ in the case of BH-BH formation. The main results presented in
this section are summarized in Table~1 (top 3 entries) and illustrated in 
Figure~1 (top panel).

\subsection{Range of evolutionary uncertainties}

Alternatively to our standard approach, we employ WR wind mass 
loss rates adopted from Zdziarski et al. (2012):
\begin{equation}
(dM/dt) = 1.9 \times 10^{-5} \left({M_{\rm WR} \over 14.7 M_\odot}\right)^{2.93} 
\mpy.
\label{empirical}
\end{equation}
Zdziarski et al. (2012) selected data points from Nugis \& Lamers (2000) for 
WR stars of similar mass and type as the one in Cyg X-3. Nugis \& Lamers
(2000) have estimated clumping corrected mass loss rates based on
observations and modeling of 64 Galactic WR stars. We refer to this wind
mass loss prescription as empirical as it is heavily derived from observations
in contrast to the method used in our standard approach (mostly originating
from theoretical predictions). 
In particular Zdziarski et al. (2012) employed data points from Table 5 of
Nugis \& Lamers (2000) for WN stars. It is worth noting that the majority of
these WN stars ($19$ out of $34$) are in binary systems. 
The results of calculations with those winds are presented in Table~1 (models 
marked with ``Wind: empiri'' entry) and in Figure~1 (bottom panel). It is
noted that both wind mass loss prescriptions give very similar results. There
are no significant changes in our analysis. There is only a slight increase of
BH mass ($M_{\rm BH,2}=8.0\msun$) in the case of BH-BH formation.  

Next, we alternate our approach to natal kicks. We adopt high kicks
for all compact objects. It means that both neutron stars and black holes
receive the kicks from the same distribution described by a Maxwellian with
$\sigma=265$ km s$^{-1}$. There is no kick decrease factor applied in this 
case. Results for such an approach are listed in Table~1 (marked with
``Kicks: high'' entry). There is no significant changes for the BH-NS 
formation, as neutron stars in our standard model analysis were receiving 
almost full kicks (low fall back). There is a noticeable decrease in
the formation efficiency of close BH-BH systems ($f_{\rm close}=0.68$) that
leads to corresponding decrease in the Galactic merger rate 
(${\cal R}_{\rm MW}=1.06$ Myr$^{-1}$). The decrease is caused by binary
disruptions due to high natal kicks applied to black holes in this model. 
However, the formation efficiency is still significant despite the rather 
high kicks that black holes receive in this model, because the binary's 
initial orbital velocity, $\sim 800$ km s$^{-1}$, is already large compared 
with typical kicks.

Our calculations so far were based on the rapid supernova explosion engine, in which a
sufficiently massive progenitor does not explode at all, gets no natal kick and forms quite 
massive black hole. In the rapid model this is the reason for the emergence
of a mass gap between NS and BH masses (Belczynski et al. 2012a, although see 
Kreidberg et al. 2012). On the one hand, lower mass stars are subject to strong 
explosions and form 
neutron stars (as in case of $M_{\rm WR}=7.5, 10.3\msun$). On the other hand,
higher mass stars do not explode at all and form massive black holes (as for 
$M_{\rm WR}=14.2\msun$). This explains the rather sharp transition for all
models considered thus far from $M_{\rm WR}= 10.3\msun$ forming a neutron star
($M_{\rm NS}=1.7$--$1.8\msun$) to $M_{\rm WR}=14.2\msun$ forming quite massive 
black hole ($M_{\rm BH2}=7.4$--$8\msun$).

As a final variant, we have modified our approach to supernovae explosions. We now use 
the delayed explosion model of Fryer et al. (2012). This model generates 
a continuous compact-object mass spectrum (i.e., without a mass gap). 
The results for this set of calculations are listed in Table~1 (marked with
``SN: delayed'' entry).
Due to the delayed nature of the explosion engine, neutron stars are typically
more massive as proto-neutron stars have more time for accretion between the 
core bounce and actual explosion. The explosions are also
typically less energetic, since at later times, after some cooling, there is 
less energy to drive the explosion. This leads to two 
changes for BH-NS formation.  First, there is a noticeable increase in the formation efficiency 
(lower explosion energy leads to increased fall back and smaller natal 
kicks). Second, we note a significant increase in the NS mass.  In fact, for the 
most likely Cyg X-3 configuration ($M_{\rm BH1}=2.4$; $M_{\rm WR}=10.3\msun$), 
instead of forming a NS ($M_{\rm NS}=1.7$--$1.8\msun$ in all other models), we 
obtain a low-mass black hole ($M_{\rm BH2}=2.7\msun$) and thus transition 
from BH-NS to BH-BH formation. 
In the case of the Cyg X-3 high-mass configuration we predict as before the
formation of a BH-BH binary.  The black hole in this model is formed with moderate fall back 
($f_{\rm fb}=0.43$), and therefore receives a moderate natal kick.  It has significantly 
lower mass ($M_{\rm BH2}=3.9\msun$), since the majority of the ejected material
was not a subject to fall back.  As a result, the formation efficiency is significantly 
decreased ($f_{\rm close}=0.5$) relative to all other models.  Figure 2 shows the mass 
of the final compact object as a function of the initial mass of the WR star for all 
alternative models considered in this section.

\subsection{LIGO/VIRGO detection rate estimate}

In order to compute the advanced LIGO/Virgo event rates, we use the same detection 
threshold that was adopted in (Abadie et al., 2010) -- a signal-to-noise ratio 
(SNR) of $8$ in a single interferometer at the sensitivity of advanced LIGO. This 
is a simplified treatment, since actual detector sensitivity will depend on the 
network configuration, data quality, and the details of the search pipeline, but 
the uncertainties introduced by this simplifying assumption are smaller than the 
uncertainties in the merger rates.

Previous calculations in the population-synthesis literature (e.g., Belczynski et 
al., 2010b) were based on a simple scaling of the detection volume for arbitrary 
systems with the horizon distance $d_0$ for a NS-NS binary (the distance at which 
an optimally oriented source with component masses $M_1=M_2=1.4 \msun$  could be 
detected at an SNR of 8):
\begin{equation}
{\cal R}_{\rm LIGO} = \rho_{\rm gal} { 4 \pi \over 3} \left({d_0 \over
f_{\rm pos}}
\right)^3 \left( {\cal M}_{\rm c,dco} \over {\cal M}_{\rm c,nsns}
\right)^{15/6} {\cal R}_{\rm MW}. 
\label{Rsimple}
\end{equation}
Here, $\rho_{\rm gal}$ is the density of Milky Way-like galaxies, the chirp mass 
${\cal M}_{\rm c,nsns} \equiv M_1^{3/5} M_2^{3/5} (M_1+M_2)^{-1/5} = 1.2 \msun$ 
and the correction factor $f_{\rm pos}=2.26$ takes into account the non-uniform 
pattern of detector sensitivity and random sky location and orientation of 
sources (Finn, 1996).

However, this simple calculation suffered from several shortcomings. The sensitivity 
distance scales as $d \sim d_0({\cal M}_{\rm c,dco}/{\cal M}_{\rm c,nsns})^{5/6}$ 
only for masses sufficiently low that the gravitational-wave signal spans the bandwidth 
of the detector.  The waveforms used to compute the horizon distance included only the 
contribution from the inspiral portion, not the merger and ringdown signals.  Finally, 
cosmological effects due to the expansion of the universe have not been included: the 
redshifting (dilation) of masses in the gravitational waveform; the difference between 
the volume as a function of luminosity distance $d$ and comoving volume; and the 
difference in the rate of clocks in the source and merger frames.   Here, we include 
these effects by using the following procedure when integrating over shells centered 
on the detector:

\begin{itemize}

\item{}For each shell at a given redshift $z$, compute the luminosity distance $d(z)$ 
and the comoving volume of the shell $dV_c(z)$ using standard cosmology (see, e.g., 
Hogg, 1999) with $\Omega_M=0.272$, $\Omega_\Lambda=0.728$, $h=70.4$ (WMAP 7 results).

\item{}Given a particular combination of compact-object masses from Table 1, compute 
the SNR for an optimally oriented source at distance $d(z)$,
\begin{equation}
{\mathrm SNR}^2 = 4 \int_0^\infty \frac{|\tilde{h}(f)|^2}{S_n(f)} df,
\end{equation}
where  we use the inspiral-merger-ringdown waveform family IMRPhenomB (Ajith, 2010) 
to compute the frequency-domain waveforms $\tilde{h}(f)$ using redshifted component 
masses $M \to M (1+z)$ and the zero-detuning, high-power Advanced LIGO noise power 
spectral density $S_n(f)$.

\item{} Compute the fraction of detectable mergers $f_{\rm det}(z)$ from this shell 
by considering the fraction of sources for which the projection function 
$\Theta (\iota, \theta, \psi, \phi)$ [given in Eqs.~(3.4.b--d) of (Finn, 1996)] 
is large enough so that $(\Theta/4) {\mathrm SNR} \ge 8$.  
The cumulative distribution function of $\Theta$ can be computed numerically (the 
approximate expression given by Finn, 1996, not being sufficiently accurate) via a 
Monte Carlo over the inclination angle $\iota$ (whose cosine is uniform in $[0,1]$), 
the sky-location spherical coordinates $\theta$ (whose cosine is uniform in $[0,1]$) 
and $\phi$ (uniform in $[0, 2\pi]$), and polarization $\psi$ (uniform in $[0,\pi]$).

\item{} The contribution of the shell to the detection rate is given by 
$f_{\rm det}(z) dV_c(z) \rho_{\rm gal} {\cal R}_{\rm MW} (1+z)^{-1}$.   
We assume that the intrinsic merger rate in the source frame is independent of 
redshift (i.e., we don't include the redshift dependence of star formation rate 
or metallicity).  Thus, we use ${\cal R}_{\rm MW}$ from Table 1, and scale it by 
the space density of Milky Way-like galaxies, for which we use 0.01 per Mpc$^{3}$ 
of comoving volume.  The final factor of $1/(1+z)$ reflects the time dilation 
between the source clock, used to measure the merger rate, and the clock on Earth, 
used to measure the detection rate.

\end{itemize}

We can now integrate over multiple shells (or, in practice, compute an approximate Riemann 
sum over the shells numerically) to obtain the advanced LIGO/Virgo detection rates that we 
report in Table 1.  Including merger and ringdown waveform phases and accounting for cosmological redshift corrections decreases the detected event rates by $\gtrsim 15\%$ for BH-NS systems and by $\gtrsim 30\%$ for BH-BH systems relative to the simple scaling of Eq.~(\ref{Rsimple}), largely because the comoving volume within a luminosity distance $d$ is significantly smaller than $(4/3) \pi d^3$ at non-trivial redshifts.

\subsection{Statistical uncertainty on merger and detection rates}

We note that the rates derived above do not account for the statistical uncertainty 
associated with observing a single binary like Cygnus X-3.  In practice, we do not 
know the exact rate from a single observation even if $f_{\rm close}$ and $t_{\rm WR}$ 
are known perfectly in Eq.~\ref{eq:rate}.  Assuming that the birth of binaries like 
Cygnus X-3 is a stochastic Poisson process with rate ${\cal R}_{\rm birth}$, the 
probability of electromagnetically observing exactly one such system is
\begin{equation}
p(\rm{1\ obs.} | {\cal R}_{\rm birth} )  = {\cal R}_{\rm birth} t_{\rm WR} \exp{(-{\cal R}_{\rm birth} t_{\rm WR})}.
\end{equation} 
We can compute the probability distribution on the rate given a single observation 
by using Bayes' theorem:
\begin{equation}
p({\cal R}_{\rm birth} | \rm{1\ obs.}) \propto p({\cal R}_{\rm birth}) p(\rm{1\ obs.} | {\cal R}_{\rm birth} ),
\end{equation}
where $p({\cal R}_{\rm birth})$ is the prior probability on the birth rate of such systems.  

If a flat prior is chosen, $p({\cal R}_{\rm birth}) = {\rm const}$, the most likely birth 
rate is $1/t_{\rm WR}$, the value used in Eq.~\ref{eq:rate}.  However, the 90\% credible 
interval on the birth rate extends from $0.35/t_{\rm WR}$ (at the 5th percentile) to 
$4.74/t_{\rm WR}$ (at the 95th percentile).  

Meanwhile, if an uninformative Jeffreys prior on the rate is chosen, 
$p({\cal R}_{\rm birth}) \propto {\cal R}_{\rm birth}^{-1/2}$, the most likely birth rate 
is halved to $1/2/t_{\rm WR}$. The 90\% credible interval is shifted downward to 
$0.17/t_{\rm WR}$ -- $3.9/t_{\rm WR}$.   

Finally, we can consider the case where we assume that there is \emph{at least} one rather 
than \emph{exactly} one binary in the Galaxy in the same stage as Cygnus X-3 (i.e., there 
may be other similar Galactic systems which have not been observed).  In that case, we are 
interested in $p({\cal R}_{\rm birth} | \rm{\ge 1\ detection})$, which scales with 
$p(\rm{\ge 1\ detection} | {\cal R}_{\rm birth} ) = 1 - \exp{(-{\cal R}_{\rm birth} t_{\rm WR})}$.  
For the Jeffreys prior, the posterior birth rate distribution peaks at $1.26/t_{\rm WR}$.

The merger rate in all cases is given by ${\cal R}_{\rm MW}={\cal R}_{\rm birth}/f_{\rm close}$, 
and so ranges by the same pre-factors relative to the rate given in Eq.~\ref{eq:rate}.  The 
large range of purely statistical uncertainty, which spans about a factor of $5$ above and 
below the value in Eq.~\ref{eq:rate}, reflects the difficulty of making robust inference from 
a single observation.

\section{Discussion}

We have calculated the future evolution of Galactic binary Cyg X-3 harboring a
compact object and a WR star. We have employed a recent Cyg X-3 component mass 
estimate from Zdziarski et al. (2012) and following their arguments we have 
assumed that the compact object in Cyg X-3 is a black hole. 

Our results indicate that the future evolution and fate of Cyg X-3 is a strong 
function of mass of the WR star. Within the measurement uncertainties Cyg X-3 may
either form a close BH-BH binary at the high end of the allowed WR mass 
($M_{\rm WR} \sim 14\msun$), or form a close BH-NS ($\sim 15\%$), a wide BH-NS 
system ($\sim 15\%$) or get disrupted producing single BH and NS ($\sim 70\%$)
at the low end and middle of the allowed mass range for the WR star 
($M_{\rm WR} \sim 7$--$10\msun$).   

We have estimated the advanced LIGO/VIRGO detection rates in case of the
close BH-BH and BH-NS formation. The rates are significant: in the case of
BH-BH formation ${\cal R}_{\rm LIGO} \sim 10$ yr$^{-1}$. This is the first 
empirical estimate of a BH-BH detection rate based on a
Galactic system. Previous empirical estimates were based on extragalactic 
high mass X-ray binaries in small star forming galaxies IC10 and NGC300 
(Bulik et al. 2011). The predicted rates that are based on observations
within Milky Way (high metallicity) are much lower than estimated for the
above two low-metallicity galaxies. The BH-BH detection rate extrapolated from IC10 X-1
and NGC300 X-1 for advanced LIGO/Virgo is ${\cal R}_{\rm LIGO} \sim 2000$
yr$^{-1}$, where we used conservative mass estimates and mean merger rates 
from (Bulik et al. 2011) and converted merger rates to detection rates as 
discussed in section 2.3.
The low metallicity environment can significantly boost close BH-BH 
formation rates as explained by Belczynski et al. (2010b). 
 
The detection rate for BH-NS systems forming via the Cyg X-3 channel are low, 
${\cal R}_{\rm LIGO} \sim 0.1$ yr$^{-1}$. However, our rates are only 
{\em lower limits} as more binaries similar to Cyg
X-3 may currently exist in the local Universe and close BH-NS systems may 
potentially form via other formation channels. Among about $\sim 200$ 
Galactic and extra-galactic high-mass X-ray binaries only a handful have 
established parameters (Liu, van Paradijs \& van den Heuvel 2005, 2006) 
allowing for detection rate prediction (Belczynski et al. 2012b).
Recent population synthesis calculations provide rates that are typically 
a few detections per year for BH-NS systems for advanced detectors 
(Dominik et al. 2012; Belczynski et al. 2012c). 
Our prediction is only the second empirical estimate for BH-NS detection 
rates. The first one was obtained for another Galactic system Cyg X-1 
(Belczynski et al. 2011). We note that our current rate ($\sim 1$ detection
per decade) is $10$ times higher than the one obtained for Cyg X-1 ($\sim 1$
detection per century). 

Beyond our standard evolutionary calculations we have performed several
models to check the validity of our conclusions. We have varied WR wind
mass loss rates, natal kicks that compact objects receive in supernovae, and the
supernova explosion mechanism that alters the NS/BH mass spectrum. 
As long as we stay within the framework of rapid supernova explosion
mechanism there are no significant changes to our conclusions.  We find a range of
detection rates  ${\cal R}_{\rm LIGO}=0.09$--$0.15$ yr$^{-1}$ for 
BH-NS systems and ${\cal R}_{\rm LIGO}=7.7$--$12.4$ yr$^{-1}$ for BH-BH systems. 
Our standard supernova model, which employs rapid explosions, is motivated
by the existence of the mass gap between neutron stars and black holes  
(e.g., Bailyn 1998; Ozel et al. 2010; Farr et al. 2011). 
However, if the mass gap is not an intrinsic signature of the BH/NS mass spectrum
but is caused by some observational bias as recently claimed by Kreidberg et
al. (2012), our conclusions change. For the delayed supernova model, which is
consistent with the absence of a mass gap, we find that BH-NS formation occurs only at
the lowest allowed mass for the WR star in Cyg X-3 ($M_{\rm WR} \sim 7\msun$)
with a slightly higher detection rate ${\cal R}_{\rm LIGO}=0.17$ yr$^{-1}$.  BH-BH formation
is found in a broader mass range allowed for the WR star 
($M_{\rm WR} \sim 10$--$14\msun$).  The second BH falls right within the mass
gap ($M_{\rm BH2}=2.7$--$3.9\msun$) and the rates are significantly smaller, 
${\cal R}_{\rm LIGO}=0.41$--$2.8$ yr$^{-1}$.  Finally, we note that all merger 
and detection rates have statistical uncertainties of approximately a factor of 
$5$ in either direction due to the limited observational sample of a single 
system.

\acknowledgements
Authors acknowledge support from MSHE grant N203 404939 (KB) and NASA Grant 
NNX09AV06A to the UTB Center for Gravitational Wave Astronomy (KB). AAZ 
acknowledges support from the Polish NCN grant N N203 581240 and TB support 
from 623/N-VIRGO/09/2010/0. KB, IM and BS acknowledge the hospitality of KITP, 
supported in part by the National Science Foundation under No.~NSF Grant 
PHY11-25915.

\begin{deluxetable}{lccrcccc}
\tablewidth{520pt}
\tablecaption{The Fate of Cyg X-3}
\tablehead{
$M_{\rm BH}+M_{\rm WR}$  & Wind\tablenotemark{a}/Kicks\tablenotemark{b}/SN\tablenotemark{c} & 
Outcome\tablenotemark{d} & $f_{\rm close}$\tablenotemark{e} & $M_{\rm c,dco}$ & 
$t_{\rm WR}$ & ${\cal R}_{\rm MW}$ & ${\cal R}_{\rm LIGO}$ }
 
\startdata

$2.0 + 7.5\msun$  & theory/low/rapid & BH-NS ($2.0+1.5\msun$) & 0.18 & $1.5\msun$ & 0.95 Myr & 0.19 Myr$^{-1}$ & 0.09 yr$^{-1}$\\ 

$2.4 + 10.3\msun$ & theory/low/rapid & BH-NS ($2.4+1.7\msun$) & 0.15 & $1.8\msun$ & 0.77 Myr & 0.20 Myr$^{-1}$ & 0.13 yr$^{-1}$\\ 

$4.5 + 14.2\msun$ & theory/low/rapid & BH-BH ($4.5+7.4\msun$) & 1.00 & $5.0\msun$ & 0.64 Myr & 1.56 Myr$^{-1}$ & 11.3 yr$^{-1}$\\

                  &                           &      &            &          &                 &               \\

$2.0 + 7.5\msun$  & empiri/low/rapid & BH-NS ($2.0+1.5\msun$) & 0.18 & $1.5\msun$ & 0.95 Myr & 0.19 Myr$^{-1}$ & 0.09 yr$^{-1}$\\ 

$2.4 + 10.3\msun$ & empiri/low/rapid & BH-NS ($2.4+1.8\msun$) & 0.15 & $1.8\msun$ & 0.76 Myr & 0.20 Myr$^{-1}$ & 0.14 yr$^{-1}$\\ 

$4.5 + 14.2\msun$ & empiri/low/rapid & BH-BH ($4.5+8.0\msun$) & 1.00 & $5.2\msun$ & 0.64 Myr & 1.57 Myr$^{-1}$ & 12.4 yr$^{-1}$\\

                  &                          &      &            &          &                 &               \\

$2.0 + 7.5\msun$  & theory/high/rapid & BH-NS ($2.0+1.5\msun$) & 0.19 & $1.5\msun$ & 0.95 Myr & 0.20 Myr$^{-1}$ & 0.09 yr$^{-1}$\\ 

$2.4 + 10.3\msun$ & theory/high/rapid & BH-NS ($2.4+1.7\msun$) & 0.17 & $1.8\msun$ & 0.77 Myr & 0.22 Myr$^{-1}$ & 0.15 yr$^{-1}$\\ 

$4.5 + 14.2\msun$ & theory/high/rapid & BH-BH ($4.5+7.4\msun$) & 0.68 & $5.0\msun$ & 0.64 Myr & 1.06 Myr$^{-1}$ & 7.7 yr$^{-1}$\\

                  &                              &      &            &          &                 &               \\

$2.0 + 7.5\msun$  & theory/low/delayed & BH-NS ($2.0+1.9\msun$) & 0.26 & $1.7\msun$ & 0.95 Myr & 0.27 Myr$^{-1}$ & 0.17 yr$^{-1}$\\ 

$2.4 + 10.3\msun$ & theory/low/delayed & BH-BH ($2.4+2.7\msun$) & 0.28 & $2.2\msun$ & 0.77 Myr & 0.36 Myr$^{-1}$ & 0.41 yr$^{-1}$\\ 

$4.5 + 14.2\msun$ & theory/low/delayed & BH-BH ($4.5+3.9\msun$) & 0.50 & $3.6\msun$ & 0.64 Myr & 0.78 Myr$^{-1}$ & 2.8 yr$^{-1}$\\

\enddata
\label{data}
\tablenotetext{a}{
Either theoretically (eq.~\ref{theory}) or empirically (eq.~\ref{empirical}) based
WR wind mass loss rates are applied.} 
\tablenotetext{b}{
High: NS and BH kicks are taken from a Maxwellian with $\sigma=265$ km s$^{-1}$.
Low: the high kicks are decreased proportionally to the amount of fall back for both 
NSs and BHs.}
\tablenotetext{c}{
Compact-object formation model: either via rapid supernovae (mass gap) or delayed
supernovae (no mass gap).
}
\tablenotetext{d}{
Type of binary formed followed by the mass of compact objects.} 
\tablenotetext{e}{
We only count binaries that have formed with merger time shorter than $10$
Gyr (others are disrupted or form wider systems).}
\end{deluxetable}

\begin{figure}
\includegraphics[width=0.9\columnwidth]{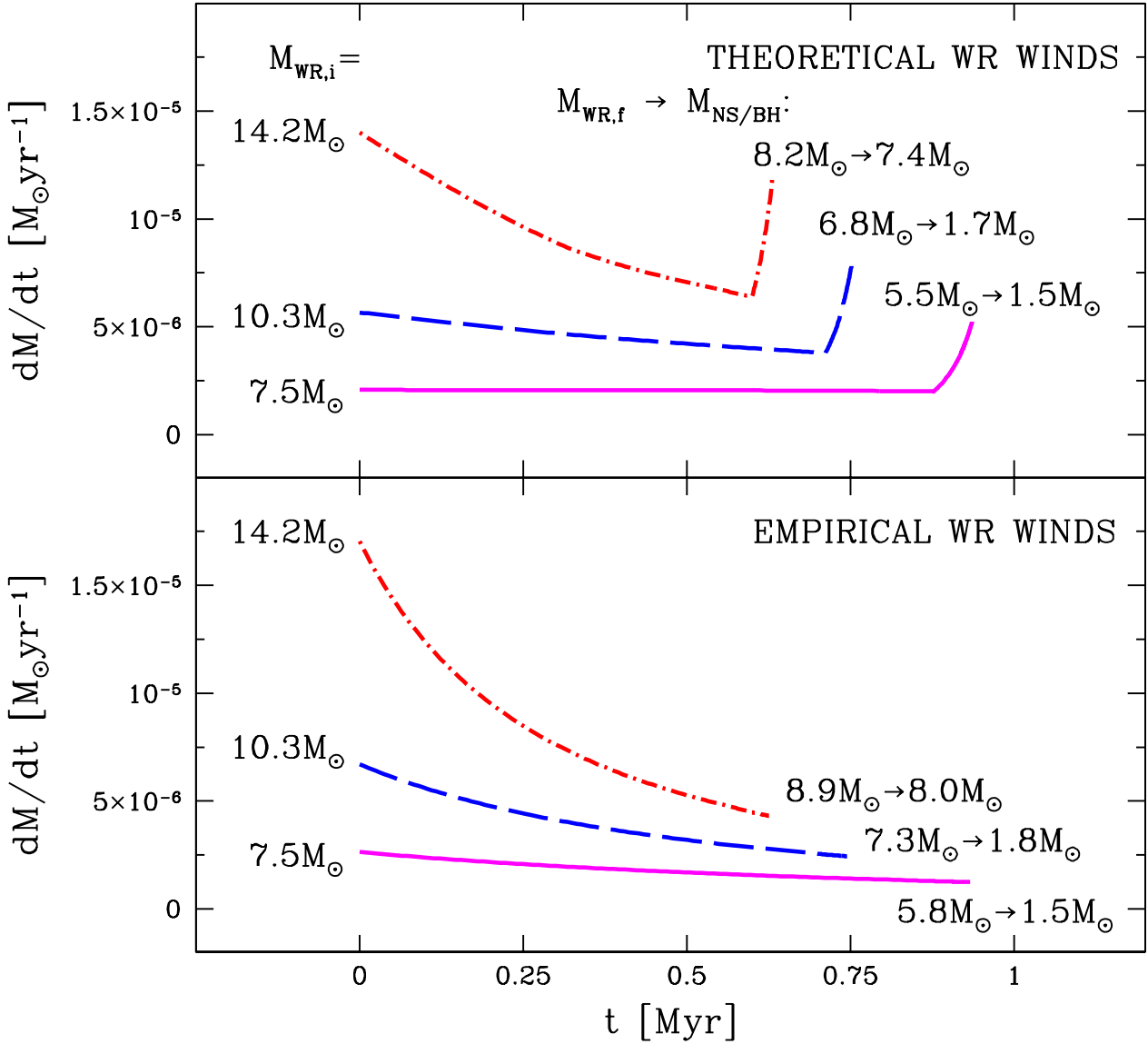}
\vspace*{0.0cm}
\caption{
Evolutionary prediction for Cyg X-3 WR star ($M_{\rm
WR,i}=10.3^{+3.9}_{-2.8}\msun$). Top panel shows the wind
mass loss rate based on theoretical calculations, while results in bottom 
panel are based on observationally estimated WR mass loss rates. Independent
of the adopted wind mass loss rate the WR component becomes either a neutron
star for low- to medium-mass progenitors ($M_{\rm WR,i}=7.5$--$10.3\msun$), and 
a black hole at the high end ($M_{\rm WR,i}=14.2\msun$) of the 
allowed WR mass range. The mass of the WR component at the end of its evolution 
is marked with $M_{\rm WR,f}$, while 
the mass of compact object formed after core collapse/supernova explosion
is marked with $M_{\rm NS/BH}$. Compact
objects with $M_{\rm NS/BH}<2\msun$ are assumed to be neutron stars, and
above that black holes.} 
\label{hel}
\end{figure}

\begin{figure}
\includegraphics[width=0.9\columnwidth]{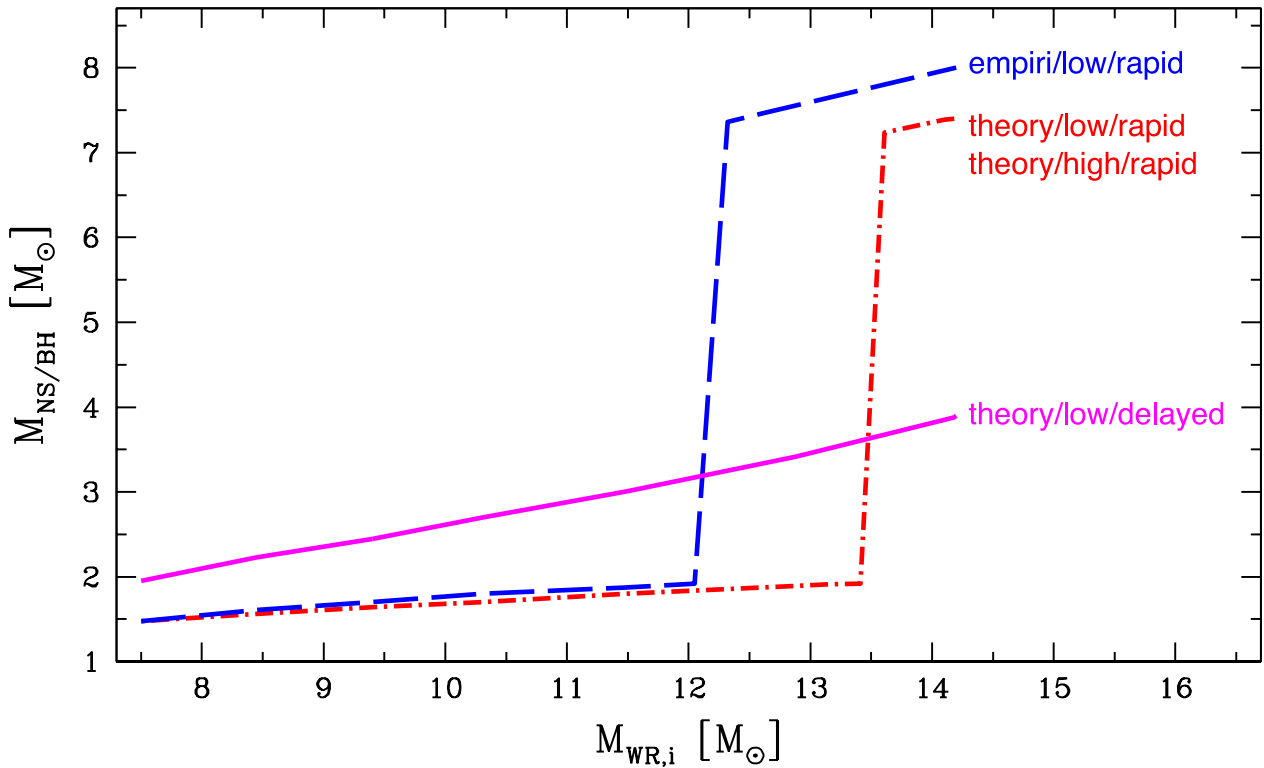}
\vspace*{0.0cm}
\caption{
The dependence of the second compact object mass on the current WR star mass 
for Cyg X-3.  There is a sharp transition from NS to BH formation for 
the rapid SN engine at around $M_{\rm WR,i} \sim 12$--$13 \msun$.  The 
delayed SN engine allows for a steady increase of compact-object mass 
and the NS/BH transition depends sensitively on the unknown maximum NS 
mass. The assumed kick velocity has no impact on the compact-object mass, 
so two kick models (theory/low/rapid and theory/high/rapid) yield the 
same mass spectrum for the second compact object.
}
\label{fmass}
\end{figure}

\end{document}